\documentclass[12pt,twoside]{article}

\usepackage{amsmath,amsthm,amssymb,euscript,amscd,graphicx, color, amsfonts, fleqn }

\usepackage{graphicx}

\newtheorem{lemma}{Lemma}[section]
\newtheorem{theorem}{Theorem}[section]

\definecolor{mybackgroundcolor}{rgb}{1,0,1}
\definecolor{plum}{rgb}{.5,0,.5}
\definecolor{dkred}{rgb}{.5,0,0}
\definecolor{ddkred}{rgb}{.35,0,0}
\definecolor{dkblue}{rgb}{.1,0,.6}
\definecolor{ddkblue}{rgb}{0,0,.25}
\definecolor{dkgreen}{rgb}{0,.5,0}
\definecolor{ddkgreen}{rgb}{0,.35,0}
\definecolor{dkgr2}{rgb}{0,.57,0}
\definecolor{dkgr3}{rgb}{0,.64,0}
\definecolor{dkgr4}{rgb}{0,.71,0}

\pagecolor{white}

\newcommand{\be}{\begin{equation}}
\newcommand{\ee}{\end{equation}}
\newcommand{\bd}{\begin{displaymath}}
\newcommand{\ed}{\end{displaymath}}
\newcommand{\bea}{\begin{eqnarray}}
\newcommand{\eea}{\end{eqnarray}}
\newcommand{\rhp}{\mathbb{C}_{+}}
\newcommand{\R}{\mathbb{R}}

\newcommand{\Hi} {{\cal H}^\infty}
\newcommand{\RHi}{{\cal RH}^\infty}

\author{Suat G\"um\"u\c{s}soy\thanks{Collaborative Center of Control Science, Department of Electrical Engineering, The Ohio State University, 2015 Neil Avenue,
Columbus, OH 43210} and Hitay \"Ozbay\thanks{Collaborative Center
of Control Science, Department of Electrical Engineering, The Ohio
State University, 2015 Neil Avenue, Columbus, OH $43210$}
\thanks{Part of this work was done at Bilkent University,
Department of Electrical and Electronics Engineering, Bilkent,
Ankara $06800$, Turkey}}
\title{On Stable $\Hi$ Controllers for Time-Delay Systems\thanks{This work was supported in part by the National Science Foundation under grant
ANI-$0073725$}}

\begin{document}
\date{}
\maketitle
\begin{abstract}

In this paper, we study the stability of suboptimal $\Hi$
controllers for time-delay systems. The optimal $\Hi$ controller
may have finitely or infinitely many unstable poles. A stable
suboptimal $\Hi$ controller design procedure is given for each of
these cases. The design methods are illustrated with examples.

\end{abstract}

\section{Introduction}
\setcounter{equation}{0}

A strongly stabilizing controller is a stable controller in a
stable feedback, \cite{V85}. In many practical applications,
strongly stabilizing controllers are desired, see e.g.
\cite{Sideris,Barabanov,Jacobus,Ito,Ganesh,ZO99,ZO00,Zhou01} and
their references. In these papers, direct design methods are given
for $\Hi$ strong stabilization for finite dimensional plant case.
The necessary and sufficient condition for strong stabilization,
parity interlacing property, is shown in \cite{Abedor89} for
single input single output delay systems. A design method to find
strongly stabilizing controller for single input single output
systems with time delays is given \cite{Suyama91} in which the
stable controller is constructed by using the unit satisfying some
interpolation conditions.

An indirect approach to design stable controller achieving  a
desired $\Hi$ performance level for time delay systems is given in
\cite{Gumussoy02}. This approach is based on stabilization of
$\Hi$ controller by another $\Hi$ controller in the feedback loop.
In \cite{Gumussoy02}, stabilization is achieved and the
sensitivity deviation is minimized. There are two main drawbacks
of this method. First, the solution of sensitivity deviation
brings conservatism because of finite dimensional approximation of
the infinite dimensional weight. Second, the stability of overall
sensitivity function is not guaranteed. Also, overall system does
not achieve the exact performance level, since the optimal $\Hi$
controller is perturbed by deviation.

Our paper focuses on strong stabilization problem for infinite
dimensional plants such that the stable controller achieves the
pre-specified suboptimal $\Hi$ performance level. When the optimal
controller is unstable (with infinitely or finitely many unstable
poles), two methods are given based on a search algorithm to find
a stable suboptimal controller. However, both methods are
conservative. In other words, there may be a stable suboptimal
controller achieving a smaller performance level, but the designed
controller satisfies the desired overall $\Hi$ norm. The stability
of optimal and suboptimal controller is discussed and necessity
conditions are given.

It is known that a $\Hi$ controller for time-delay systems with
finitely many unstable poles can be designed by the methods in
\cite{FTZ86,ZK87,OST93,TO95}. In general, weighted sensitivity
problem results in an optimal $\Hi$ controller with infinitely
unstable modes, \cite{Flamm,Lenz}.

We assume that the plant is single input single output (SISO) and
admits the representation as in \cite{TO95}, \be
P(s)=\frac{m_n(s)N_o(s)}{m_d(s)} \ee where $m_n(s)=e^{-hs}M(s)$,
$h>0$, and $M(s)$, $m_d(s)$ are finite dimensional, inner, and
$N_o(s)$ is outer, possibly infinite dimensional. The optimal
$\Hi$ controller, $C_{opt}$, stabilizes the feedback system and
achieves the minimum $\Hi$ cost, $\gamma_{opt}$: \be
\label{eq:wsm} \gamma_{opt}=\left\| \left[\begin{array}{c}
  W_1(1+PC_{opt})^{-1} \\
  W_2PC_{opt}(1+PC_{opt})^{-1}
\end{array} \right]\right\|_\infty= \inf_{C \; stabilizing \; P} \left\| \left[\begin{array}{c}
  W_1(1+PC)^{-1} \\
  W_2PC(1+PC)^{-1}
\end{array} \right]\right\|_\infty  \ee
where $W_1$ and $W_2$ are finite dimensional weights for the mixed
sensitivity minimization problem.

In the next section, the structure of optimal and suboptimal $\Hi$
controllers will be summarized. The optimal controller with
infinitely many unstable poles case is considered in Section 3.
The conditions and a design method for stable suboptimal $\Hi$
controller is given in the same section. Similar work is done in
Section 4 for the optimal controller with finitely many unstable
poles. Examples related for these design methods are presented in
Section 5, and concluding remarks can be found in Section 6.

\section{Structure of $\Hi$ Controllers}

Assume that the problem (\ref{eq:wsm}) satisfies $(W_2 N_o), (W_2
N_o)^{-1}\in\Hi$, then optimal $\Hi$ controller can be written as,
\cite{FOT},

\be \label{eq:Copt} C_{opt}(s)=E_{\gamma_{opt}}(s)
m_d(s)\frac{N_o^{-1}(s)F_{\gamma_{opt}}(s)L(s)}{1+m_n(s)F_{\gamma_{opt}}(s)L(s)}
\ee where
$E_{\gamma}=\left(\frac{W_1(-s)W_1(s)}{\gamma^2}-1\right)$, and
for the definition of the other terms, let the right half plane
zeros of $E_\gamma(s)$ be $\beta_i$, $i=1,\ldots,n_1$, the right
half plane poles of $P(s)$ be $\alpha_i$, $i=1,\ldots,l$ and that
of $W_1(-s)$ be $\eta_i$ $i=1,\ldots,n_1$. Then,
$F_{\gamma}(s)=G_{\gamma}(s)\prod_{i=1}^{n_1}\frac{s-\eta_i}{s+\eta_i}$
where \be \label{eq:spectralfactorization}
G_{\gamma}(s)G_{\gamma}(-s)=\left(1-\left(\frac{W_2(-s)W_2(s)}{\gamma^2}-1\right)E_\gamma\right)^{-1}
\ee and $G_\gamma,G_\gamma^{-1}\in\Hi$, and
$L(s)=\frac{L_2(s)}{L_1(s)}$ , $L_1(s)$ and $L_2(s)$ are
polynomials with degrees less than or equal to $(n_1+l-1)$ and
they are determined by the following interpolation conditions,
\bea \label{eq:interpcond}
0&=&L_1(\beta_i)+m_n(\beta_i)F_{\gamma}(\beta_i)L_2(\beta_i) \quad i=1,\ldots,n_1 \\
\nonumber 0&=&L_1(\alpha_i)+m_n(\alpha_i)F_{\gamma}(\alpha_i)L_2(\alpha_i) \quad i=1,\ldots,l \\
\nonumber 0&=&L_2(-\beta_i)+m_n(\beta_i)F_{\gamma}(\beta_i)L_1(-\beta_i) \quad i=1,\ldots,n_1 \\
\nonumber
0&=&L_2(-\alpha_i)+m_n(\alpha_i)F_{\gamma}(\alpha_i)L_1(-\alpha_i)
\quad i=1,\ldots,l. \eea The optimal performance level,
$\gamma_{opt}$, is the largest $\gamma$ value such that spectral
factorization (\ref{eq:spectralfactorization}) exists and
interpolation conditions (\ref{eq:interpcond}) are satisfied.

Similarly, the suboptimal controller achieving the performance
level, $\rho$, can be defined as, \be \label{eq:Csubopt}
C_{subopt}(s)=E_{\rho}(s)
m_d(s)\frac{N_o^{-1}(s)F_{\rho}(s)L_U(s)}{1+m_n(s)F_{\rho}(s)L_U(s)}\ee
where $\rho>\gamma_{opt}$ and
$L_U(s)=\frac{L_{2U}(s)}{L_{1U}(s)}=\frac{L_2(s)+L_1(-s)U(s)}{L_1(s)+L_2(-s)U(s)}$
with $U\in\Hi$, $\|U\|_\infty\leq1$. The polynomials, $L_1(s)$ and
$L_2(s)$, have degrees less than or equal to $n_1+l$. Same
interpolation conditions are valid with $\rho$ instead of
$\gamma$. Moreover, there are two additional interpolation
conditions for $L_1(s)$ and $L_2(s)$: \bea
0&=&L_2(-a)+(E_\rho(a)+1)F_\rho(a)m_n(a)L_1(-a) \\
0&\neq&L_1(-a) \eea where $a\in\mathbb{R}^+$ is arbitrary. The
above terms and notations are the same as in \cite{FOT}.

Note that the unstable zeros of $E_{\gamma_{opt}}$ and $m_d$ are
always cancelled by the denominator in (\ref{eq:Copt}). Therefore,
$C_{opt}$ is stable if and only if the denominator in
(\ref{eq:Copt}) has no unstable zeros except the unstable zeros of
$E_{\gamma_{opt}}$ and $m_d$ (multiplicities considered). Same
conclusions are valid for the suboptimal case, $C_{subopt}$ is
stable provided that the denominator in (\ref{eq:Csubopt}) has
unstable zeros only at the unstable zeros of $E_{\rho}$ and $m_d$
(again, multiplicities considered).

It is clear that the optimal, respectively suboptimal, controllers
have infinitely many unstable poles if and only if there exists
$\sigma_o>0$ such that the following inequality holds \be
\lim_{\omega\rightarrow\infty}|F_{\gamma_{opt}}(\sigma_o+j\omega)L_{opt}(\sigma_o+j\omega)|>1,
\ee respectively, \be
\lim_{\omega\rightarrow\infty}|F_{\rho}(\sigma_o+j\omega)L_U(\sigma_o+j\omega)|>1.
\ee The controller may have infinitely many poles because of the
delay term in the denominator. All the other terms are finite
dimensional.

Even when the optimal controller has infinitely many unstable
poles, a stable suboptimal controller may be found by proper
selection of the free parameter $U(s)$. In Section 3 this case is
discussed.

Note that the previous case covers one and two block cases (i.e.,
$W_2=0$ and $W_2\neq$0 respectively). When $F_{\gamma_{opt}}$ is
strictly proper, then the optimal and suboptimal controllers may
have only finitely many unstable poles. Existence of stable
suboptimal $\Hi$ controllers and their design will be discussed in
Section 4 for this case.

\section{Stable suboptimal $\Hi$ controllers, when the optimal
controller has infinitely many unstable poles}
\label{eq:infinitelymany}

The following lemma gives the necessary condition for a suboptimal
controller to have finitely many unstable poles.

\begin{lemma} \label{eq:infitycond}

Assume that the optimal controller has infinitely many unstable
poles and $U(s)$ is finite dimensional, the suboptimal controller
has finitely many unstable poles if and only if \be
\label{eq:inftycond}
\lim_{\omega\rightarrow\infty}|F_\rho(j\omega)L_U(j\omega)|\leq1
\ee
\end{lemma}
\noindent \textbf{Proof} Assume that the suboptimal controller has
infinitely many unstable poles, then the equation\bd
1+e^{-h(\sigma+j\omega)}M(\sigma+j\omega)F_\rho(\sigma+j\omega)L_U(\sigma+j\omega)=0
\ed has infinitely many zeros in the right half plane,i.e., there
exists $\sigma=\sigma_o>0$ and for sufficiently large $\omega$,
\be \label{eq:delay1}
1+e^{-h(\sigma_o+j\omega)}\lim_{\omega\rightarrow\infty}\left(F_\rho(\sigma_o+j\omega)L_U(\sigma_o+j\omega)\right)=0
\ee will have infinitely many zeros. Since $F_\rho$ and $L_U$ are
finite dimensional, \bea
\nonumber \lim_{\omega\rightarrow\infty}F_\rho(j\omega)&=&\lim_{\omega\rightarrow\infty}F_\rho(\sigma+j\omega) \\
\nonumber
\lim_{\omega\rightarrow\infty}L_U(j\omega)&=&\lim_{\omega\rightarrow\infty}L_U(\sigma+j\omega)
\quad \forall \; \sigma>0. \eea By using this fact, we can rewrite
(\ref{eq:delay1}) as, \be
1+e^{-h(\sigma_o+j\omega)}\lim_{\omega\rightarrow\infty}\left(F_\rho(j\omega)L_U(j\omega)\right)=0
\ee which implies that in order to have infinitely many zeros, the
condition in lemma should satisfied. Conversely, a similar idea
can be used to show that (\ref{eq:inftycond}) implies finitely
many unstable poles.\qed

Note that this lemma is valid not only for only finite dimensional
$U(s)$ term, but also for any $U\in\Hi$, $\|U\|_\infty\leq1$
 provided that
\be
\lim_{\omega\rightarrow\infty}U(j\omega)=\lim_{\omega\rightarrow\infty}U(\sigma+j\omega)=u_\infty,
\quad \forall \; \sigma>0. \ee is satisfied where $u_\infty\in
\R$. Also, we can find conditions on $U$ which guarantees finitely
many unstable poles by using the lemma.

Assume that $U(s)$ is finite dimensional and bi-proper, and define
\bea \nonumber
f_\infty&=&\lim_{\omega\rightarrow\infty}|F_\rho(j\omega)|>1 \\
\nonumber u_\infty&=&\lim_{\omega\rightarrow\infty}U(j\omega) \\
\nonumber
k&=&\lim_{\omega\rightarrow\infty}\frac{L_{2}(j\omega)}{L_{1}(j\omega)}
\eea
\begin{lemma} \label{eq:inftyvalues}
The suboptimal controller has finitely many unstable poles if and
only if the following inequalities hold: \be \label{eq:firstineq}
|k|\leq\frac{1}{f_\infty}, \quad
|u_\infty|\leq\frac{1-f_\infty|k|}{f_\infty-|k|} \ee when
$(n_1+l)$ is odd (even) and $ku_\infty<0,(ku_\infty>0)$, and \be
\label{eq:secondineq} |k|<1, \quad
\frac{f_\infty|k|-1}{f_\infty-|k|}<|u_\infty|<\frac{f_\infty|k|+1}{f_\infty+|k|}
\ee when $(n_1+l)$ is odd (even) and $ku_\infty>0,(ku_\infty<0)$.
\end{lemma}
\noindent \textbf{Proof} By using Lemma \ref{eq:infitycond}, when
$(n_1+l)$ is odd (even) and $ku_\infty<0,(ku_\infty>0)$, we can
re-write (\ref{eq:inftycond}) as\bd
f_\infty\frac{|k|+|u_\infty|}{1+|k||u_\infty|}\leq1. \ed After
algebraic manipulations and using $f_\infty>1$, we can show that
(\ref{eq:firstineq}) satisfies this condition. Similarly, when
$(n_1+l)$ is odd (even) and $ku_\infty>0,(ku_\infty<0)$,
(\ref{eq:inftycond}) is equivalent to \bd
f_\infty\left|\frac{|k|-|u_\infty|}{1-|k||u_\infty|}\right|\leq1.
\ed, and (\ref{eq:secondineq}) satisfies this condition.\qed \\
Note that $u_\infty$ is a design parameter and the range can be
determined, by given $f_\infty$ and $k$.

\begin{theorem} \label{eq:thmss}
Assume that the optimal and central suboptimal controller (when
$U=0$) has infinitely many unstable poles, if there exists
$U\in\Hi$, $\|U\|_\infty<1$ such that $L_{1U}$ has no $\rhp$ zeros
and $|L_U(j\omega)F_\rho(j\omega)|\leq1$, $\forall \;
\omega\in[0,\infty)$, then the suboptimal controller is stable.
\end{theorem}
\noindent \textbf{Proof} Assume that there exists $U$ satisfying
the conditions of the theorem. By maximum modulus theorem, \bd
|1+e^{-hs_o}M(s_o)F_\rho(s_o)L_U(s_o)|>1-e^{-h\sigma}|F_\rho(j\omega)L_U(j\omega)|>0,
\ed therefore, there is no unstable zero, $s_o=\sigma+j\omega$
with $\sigma>0$. Since, all imaginary axis zeros are cancelled by
$E_\rho$, the suboptimal controller has no unstable poles.\qed

The theorem has two disadvantages. First, there is no information
for calculation of an appropriate parameter, $U$. Second, the
inequality brings conservatism and there may exist stable
suboptimal controllers even when the condition is violated. It is
difficult to reveal the first problem, therefore it is better to
use first order bi-proper function for $U$. For the second
problem, define $\omega_{max}$ and $\eta_{max}$ as, \bea \nonumber
\omega_{max}&=&
\max_{|L_U(j\omega)F_\rho(j\omega)|=1}{\omega}, \\
\nonumber
\eta_{max}&=&\max_{\omega\in[0,\infty)}{|L_U(j\omega)F_\rho(j\omega)|}.
\eea It is important to design $\omega_{max}$ and $\eta_{max}$ as
small as possible by the choice of $U$. Otherwise, at high
frequencies the delay term will generate unstable zeros when
$\omega_{max}$ is large. Similarly, when $\eta_{max}$ is large,
although $\omega_{max}$ is small, it may cause unstable zeros. The
design method given below searches for a first order $U$, and it
is based on the above ideas. An example will be given
in Section 5.\\

\noindent \textbf{Algorithm}\\
Define $U(s)=u_\infty\left(\frac{u_z+s}{u_p+s}\right)$ such that
$u_\infty,u_p,u_z\in\R$, $|u_\infty|<1$, $u_p>0$ and $u_p\geq
u_\infty|u_z|$,
\begin{description}
  \item[1)] Fix $\rho>\gamma_{opt}$,
  \item[2)] Obtain $f_\infty$ and $k$ from the central suboptimal
  controller,
  \item[3)] Calculate admissible values of $u_\infty$ by using Lemma
  (\ref{eq:inftyvalues}),
  \item[4)] Search admissible values for $(u_\infty,u_p,u_z)$ such
  that $L_{1U}(s)$ is stable,
  \item[5)] Find the minimum $\omega_{max}$ and $\eta_{max}$ for all admissible
  $(u_\infty,u_p,u_z)$.
  \item[6)] Check in the region $D=\{s=\sigma+j\omega, \sigma\geq0:
  |e^{-hs}M(s)F_\rho(s)L_U(s)|>1\}$ whether
  $1+e^{-hs}M(s)F_\rho(s)L_U(s)$ has no $\rhp$ zeros except unstable
  zeros of $E_\rho$ and $m_d$.
\end{description}

When the central suboptimal controller has infinitely many
unstable poles, it is not possible to obtain a stable suboptimal
controller by a choice of U as strictly proper or inner function.
Once we find a $U$ from the above algorithm, the resulting stable
suboptimal $\Hi$ controller can be represented as cascade and
feedback connections of  finite dimensional terms and a finite
impulse response filter that does not have unstable pole-zero
cancellations in the controller, as explained in \cite{MZ00}.

\section{Stable suboptimal $\Hi$ controllers, when the optimal
controller has finitely many unstable poles}
\label{eq:finitelymany}

In this section, we will derive the conditions for the $\Hi$
controllers to have finitely many unstable poles. A sufficient
condition for the existence of stable suboptimal $\Hi$ controllers
is given, and a design method will be derived.

The optimal and suboptimal controllers have infinitely many
unstable poles, when $F_{\gamma_{opt}}L_{opt}$ and $F_{\rho}L_{U}$
has magnitude greater than one as $\omega\rightarrow\infty$. It is
not difficult to see that controllers will have finitely many
unstable poles if $F_{\gamma_{opt}}$ and $F_\rho$ are strictly
proper. Since, these terms decrease as $\omega\rightarrow\infty$
and delay term decays as $\sigma$ increases, only finitely many
unstable poles may appear. Clearly, there may be $\Hi$ controllers
(depending on parameter values) with finitely many poles while
$F_{\gamma_{opt}}$ and $F_\rho$ are bi-proper. However, it is
important to find the sufficient conditions when they are strictly
proper, which results in controllers with finitely many unstable
poles regardless of parameters.

\begin{lemma}
The $\Hi$ controller has finitely many unstable poles if the plant
is strictly proper and $W_1$ is proper (in the sensitivity
minimization problem) and, $W_1$ is proper and $W_2$ is improper
(in the mixed sensitivity minimization problem).
\end{lemma}
\noindent \textbf{Proof} Transfer function $F(s)$can be written as
ratio of two polynomials, $N_F$ and $D_F$, with degrees $m$ and
$n$ respectively. We can define relative degree function, $\phi$,
as \bd \phi(F(s))=\phi\left(\frac{N_F(s)}{D_F(s)}\right)=n-m.\ed
Note that $\phi(F_1(s)F_2(s))=\phi(F_1(s))+\phi(F_2(s))$ and
$\phi(F(s)F(-s))=2\phi(F(s))$.

The optimal controller has finitely many unstable poles if
$F_{\gamma_{opt}}$ is strictly proper, i.e.
$\phi(F_{\gamma_{opt}}(s))>0$. To show this, we can write by using
definition of $F_{\gamma_{opt}}$ and
(\ref{eq:spectralfactorization}), \bea
\nonumber \phi(F_{\gamma_{opt}}(s))&=&\phi(G_{\gamma_{opt}}(s)), \\
\nonumber &=&\frac{1}{2}\;\phi(\left(W_1(s)W_1(-s)+W_2(s)W_2(-s)-\gamma_{opt}^{-2}W_1(s)W_1(-s)W_2(s)W_2(-s)\right)^{-1}), \\
\nonumber &=&-\frac{1}{2}\;\phi(\left(W_1(s)W_1(-s)+W_2(s)W_2(-s)-\gamma_{opt}^{-2}W_1(s)W_1(-s)W_2(s)W_2(-s)\right)), \\
\nonumber &=&-\frac{1}{2}\;\min{\left\{\phi(W_1(s)W_1(-s)), \phi(W_2(s)W_2(-s)), \phi(W_1(s)W_1(-s)W_2(s)W_2(-s))\right\}}, \\
\nonumber
&=&-\min{\left\{\phi(W_1(s)),\phi(W_2(s)),\phi(W_1(s))+\phi(W_2(s))\right\}}.
\eea Strictly properness of $F_{\gamma_{opt}}$ implies, \be
\label{eq:mininequality}
\min{\left\{\phi(W_1(s)),\phi(W_2(s)),\phi(W_1(s))+\phi(W_2(s))\right\}}<0.
\ee We know that $\phi(W_1(s))\geq0$ and $\phi(W_2(s))\leq0$,
\cite{FOT}. Therefore, the inequality (\ref{eq:mininequality}) is
satisfied if and only if $\phi(W_1(s))\geq0$ and $\phi(W_2(s))<0$
are valid which means that $W_1(s)$ is proper and $W_2(s)$ is
improper. Since we have $(W_2 N_o)^{-1}\in\RHi$ \cite{FOT}, we can
conclude that the plant is strictly proper. Same proof is valid
for the suboptimal case. \qed

We know that the suboptimal controllers are written as
(\ref{eq:Csubopt}), \bd C_{subopt}(s)=E_{\rho}(s)
m_d(s)\frac{N_o^{-1}(s)F_{\rho}(s)L_U(s)}{1+m_n(s)F_{\rho}(s)L_U(s)}\ed
we can rewrite the suboptimal controllers as, \bd
C_{subopt}(s)=\left(\frac{N_o^{-1}(s)F_\rho(s)}{dE_\rho(s)dm_d(s)}\right)\left(\frac{L_2(s)+L_1(-s)m_n(s)F_\rho(s)}{P_1(s)+P_2(s)U(s)}\right)\ed
where \bea
\nonumber P_1(s)&=&\frac{L_1(s)+L_2(s)m_n(s)F_\rho(s)}{dE_\rho(s)dm_d(s)}, \\
\nonumber
P_2(s)&=&\frac{L_2(-s)+L_1(-s)m_n(s)F_\rho(s)}{nE_\rho(s)nm_d(s)},
\eea and $nE_\rho$, $dE_\rho$ and $nm_d$, $dm_d$ are numerator and
denominator of $E_\rho$ and $m_d$ respectively. Denominators of
$P_1$ and $P_2$ are cancelled by numerators.

Note that unstable poles of $C_{subopt}$ are the zeros of
$P_1+P_2U$. If there exists a $U\in\RHi$ with $\|U\|_\infty<1$,
such that $P_1+P_2U$ has no unstable zeros, then the corresponding
suboptimal controller is stable.

Assume that $F_\rho$ is strictly proper which implies $P_1$ and
$P_2$ has finitely many unstable zeros. The suboptimal controller
is stable if and only if $S_U:=(1+\tilde{P}U)^{-1}$ is stable
where $\tilde{P}=\frac{P_2}{P_1}$. Note that since $P_1$ and $P_2$
has finitely many unstable zeros, we can write $\tilde{P}$ as, \bd
\tilde{P}=\frac{\tilde{M}}{\tilde{M_d}} \tilde{N_o} \ed where
$\tilde{M}$ and $\tilde{M_d}$ are inner, finite dimensional and
$\tilde{N_o}$ is outer and infinite dimensional. Finding stable
$S_U$ with $U\in\Hi$ is a sensitivity minimization problem with
stable controller which is considered in \cite{Ganesh}. However,
in our case, $U$ has a norm restriction as $\|U\|_\infty\leq1$ and
we can write $U$ as, \bd
U(s)=\left(\frac{1-S_U(s)}{S_U(s)}\right)\left(\frac{P_1(s)}{P_2(s)}\right).\ed

Define $\mu_{opt}$ as, \bd
\mu_{opt}=\inf_{U\in\Hi}\|S_U\|_\infty=\inf_{U\in\Hi}\|(1+\tilde{P}U)^{-1}\|_\infty.\ed
If we fix $\mu$ as $\mu>\mu_{opt}$, then there exists a free
parameter $Q$ ($Q\in\Hi$ and $\|Q\|_\infty\leq1$) which
parameterizes all functions stabilizing $S_U$ and achieving
performance level $\mu$. We will show that the sensitivity
function achieving performance level $\mu$ as $S_{U,\mu}(Q)$.

\begin{lemma} \label{eq:finitepoles}
Assume that $W_1$ and $W_2$ are proper and improper respectively.
If there exists $\mu_o>\mu_{opt}$ and $Q_o$ with $Q_o\in\Hi$ and
$\|Q_o\|_\infty\leq1$ satisfying \be
\left|\left(\frac{1-S_{U,\mu_o}(Q_o(j\omega))}{S_{U,\mu_o}(Q_o(j\omega))}\right)\left(\frac{P_1(j\omega)}{P_2(j\omega)}\right)\right|\leq1,
\ee then the suboptimal controller, $C_{subopt}$, achieves the
performance level $\rho$ by selecting the parameter $U$ as, \be
U(s)=\left(\frac{1-S_{U,\mu_o}(Q_o(s))((s)}{S_{U,\mu_o}(Q_o(s))}\right)\left(\frac{P_1(s)}{P_2(s)}\right)
\ee
\end{lemma}
\noindent \textbf{Proof} The result of theorem is immediate. Since
$Q_o$ satisfies the norm condition of $U$ and makes
$S_{U,\mu}(Q_o)$ stable, the suboptimal controller has no right
half plane poles by selection of $U$ as shown in theorem.\qed

A stable suboptimal controller can be designed by finding $Q_o$
for $\mu_o$. By using a search algorithm, we can find $Q_o$
satisfying the norm condition for $U$. Instead of finding $U$
resulting stable suboptimal controller, the problem is converted
finding $Q_o$ satisfying the norm condition. First problem needs
to check whether a quasi-polynomial has unstable zeros. However,
by using the theorem, this problem reduced into searching stable
function with infinity norm less than one and satisfying norm
condition for $U$. Conservatively, the search algorithm for $Q_o$
can be done for first order bi-proper functions such that
$Q_o(s)=u_\infty\left(\frac{s+z_u}{s+p_u}\right)$ where $p_u>0$,
$z_u\in\R$, and $|u_\infty|\leq\max{\{1,\frac{p_u}{|z_u|}\}}$. The
algorithm for this approach is explained below.

\noindent \textbf{Algorithm}\\
Assume that the optimal and central suboptimal controllers have
finitely many unstable poles. We can design a stable suboptimal
$\Hi$ controller by using the following algorithm:

\begin{description}
  \item[1)] Fix $\rho>\gamma_{opt}$,
  \item[2)] Obtain $P_1$ and $P_2$. If $P_1$ has no unstable
  zeros, then the suboptimal controller is stable for $U=0$. If
  not, go to step 3.
  \item[3)] Define the right half plane zeros of $P_1$ and $P_2$
  as $p_i$ and $s_i$ respectively. Note that these are right half plane zeros of
  $\tilde{M}_d(s)$ and $\tilde{M}(s)$ respectively. Calculate
  $w_i=\frac{1}{\tilde{M_d}(s_i)}$ and $z_i=\frac{s_i-a}{s+a}$
  where $a>0$.
  \item[4)] Search for minimum $\mu$ which makes the Pick matrix
  positive semi-definite,\be
  Q_P\{\mu\}_{(i,k)}=\left(\frac{-\ln\frac{w_i}{\mu}-\ln\frac{\bar{w}_k}{\mu}+j2\pi(n_k-n_i)}{1-z_i\bar{z}_k}\right)
  \ee
  where $n_{[.]}$ is integer. Note that most of the integers will
  not result in positive semi-definite Pick matrix. Therefore, for
  each integer set, we can find the smallest $\mu$ and $\mu_{opt}$
  will be the minimum of these vales. For details, see
  \cite{Ganesh}.
  \item[5)] After the integer set and $\mu_{opt}$ is found, the function
  $g(z)\in\Hi$ can be obtained satisfying interpolation conditions,\be
  g(z_i)=-\ln{\frac{w_i}{\mu_{opt}}}-j2\pi n_i \ee by
  Nevanlinna-Pick interpolation approach \cite{FOT},\cite{ZO98}. Then, we can
  write $S_U(s)=\mu_{opt}\tilde{M}_d(s)e^{-G(s)}$ where
  $G(s)=g(\frac{s-a}{s+a})$ and obtain $U(s)$. Check the norm condition
  $\|U\|_\infty\leq1$. If it is satisfied, then, $U(s)$ results in
  stable suboptimal controller achieving performance level $\rho$.
  If not, go to next step.
  \item[6)] Increase $\mu$ such that $\mu>\mu_{opt}$. For all
  possible integer set, obtain $g(z)\in\Hi$ with interpolation
  conditions, \be
  g(z_i)=-\ln{\frac{w_i}{\mu}}-j2\pi n_i. \ee Note that since
  $g(z)$ has a free parameter $q(z)$ ($q\in\Hi$ and $\|q\|_\infty\leq1$), we can write the function as
  $g(z,q)$. Then, search for parameters ($u_\infty$,$z_u$,$p_u$)
  satisfying \be \left|\left(\frac{1-\mu \tilde{M}_d(j\omega)e^{-G(j\omega,Q)}}{\mu \tilde{M}_d(j\omega)e^{-G(j\omega,Q)}}\right)\left(\frac{P_1(j\omega)}{P_2(j\omega)}\right)\right|\leq1, \quad \forall
  \omega\in[0,\infty)\ee where
  $G(s,Q(s))=g(\frac{s-a}{s+a},q(\frac{s-a}{s+a}))$ and
  $Q(s)=u_\infty\left(\frac{s+z_u}{s+p_u}\right)$ as defined
  before. If one of the parameter set satisfies the inequality, then $Q_o=u_{\infty,o}\left(\frac{s+z_{u,o}}{s+p_{u,o}}\right)$
  and corresponding $U$ results in a stable suboptimal $\Hi$ controller. If no parameter set satisfies the inequality, go
  to step 6, and repeat the procedure for sufficiently high $\mu$,
  until a pre-specified maximum is reached, in which case go next
  step.
  \item[7)] Increase $\rho$, go to step 2, if a maximum pre-specified $\rho$ is reached, stop. This method fails to provide a stable $\Hi$ controller.
\end{description}

An illustrative example is presented in Section 5.

\section{Examples}

Two examples will be given in this section. In the first example,
the optimal and central suboptimal controllers have infinitely
many unstable poles; by using the design method, we show that
there exists a stable suboptimal controller even the magnitude
condition ($|L_U(j\omega)F_\rho(j\omega)|\leq1$) is violated for
low frequencies. In other words,  the example illustrates that the
conditions in (\ref{eq:thmss}) are sufficient.

The second example explains the design method for stable
suboptimal $\Hi$ controller whose central controller is unstable
with finitely many unstable poles and implements the algorithm
step by step as mentioned in section \ref{eq:finitelymany}.

\subsection{Example}
Let $P(s)=e^{-0.1s}\left(\frac{s-1}{s+1}\right)$ and choose
$W_1(s)=\frac{1+0.6s}{s+1}$ and $W_2=0$ (one-block problem). Using
Skew-Teoplitz approach in \cite{FOT}, the minimum $\Hi$ value,
$\gamma_{opt}$, is $0.8108$. The optimal controller has infinitely
many unstable poles converging to $s=3.0109\pm
j\frac{(2k+1)\pi}{h}$ as $k\rightarrow\infty$. If central
suboptimal controller ($U=0$) is calculated for $\rho=0.814$, it
has infinitely many unstable poles converging to $s=2.445\pm
j\frac{(2k+1)\pi}{h}$ as $k\rightarrow\infty$. The suboptimal
controllers can be represented as, \be C_{subopt}(s)=E_{\rho}(s)
\frac{F_{\rho}(s)L_U(s)}{1+m_n(s)F_{\rho}(s)L_U(s)}\ee where \bea
\nonumber m_n(s)&=&e^{-0.1s}\left(\frac{s-1}{s+1}\right), \\
\nonumber E_\rho(s)&=&\frac{0.3374+0.3026s^2}{0.6626(1-s^2)},\\
\nonumber F_\rho(s)&=&0.814\left(\frac{1-s}{1+0.6s}\right), \\
\nonumber
L_U(s)&=&\frac{L_{2U}(s)}{L_{1U}(s)}=\frac{L_2(s)+L_1(-s)U(s)}{L_1(s)+L_2(-s)U(s)},\\
\nonumber L_2(s)&=&-(0.9413s+1.8716),\\
\nonumber L_1(s)&=&(s+1.8373). \eea

\begin{figure}
\begin{minipage}[b]{.5\linewidth}
\centering \includegraphics[width=3in]{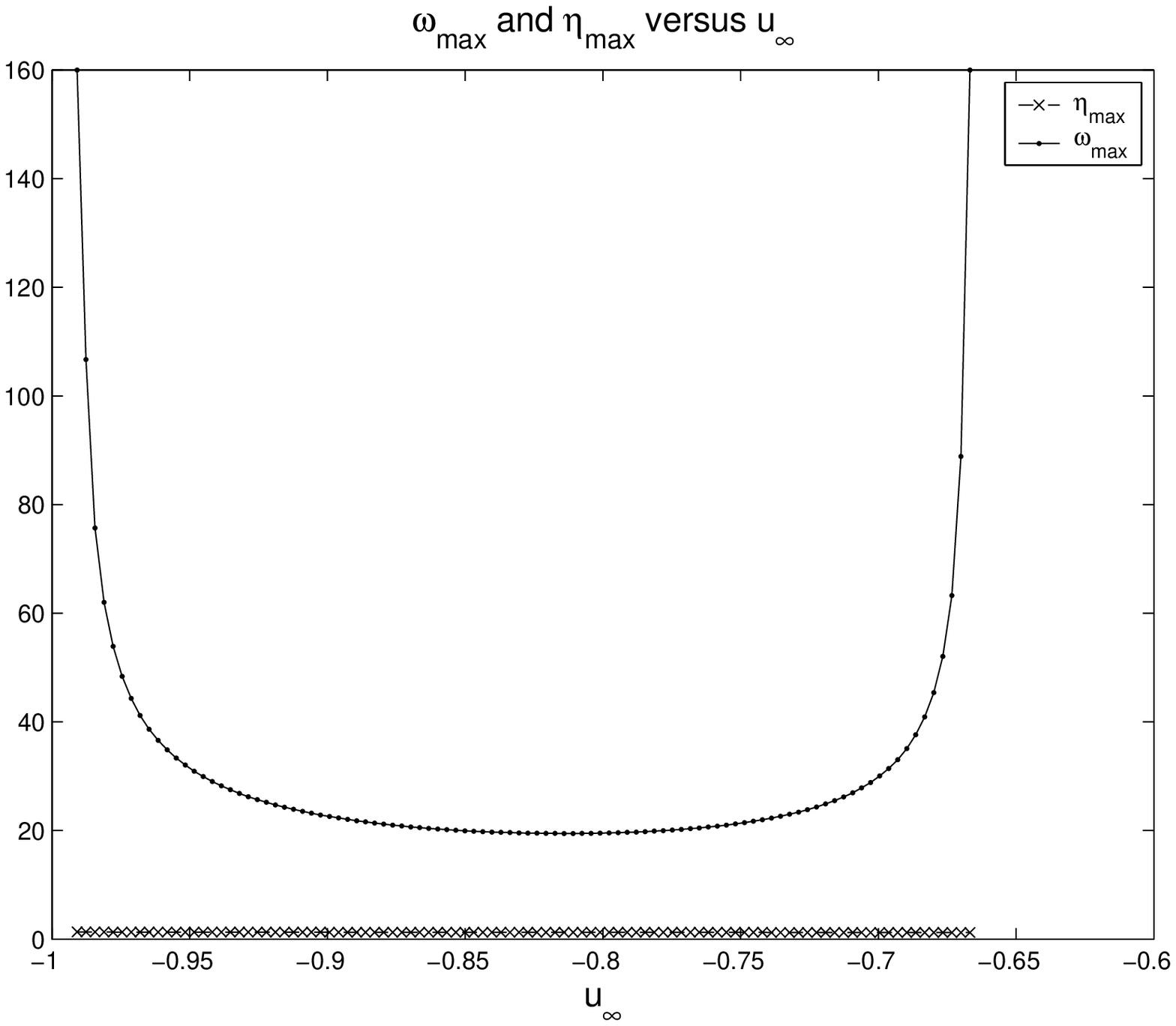}
\caption{$w_{max}$ and $\eta_{max}$ versus $u_\infty$}
\label{fig:wetamax}
\end{minipage}%
\begin{minipage}[b]{.5\linewidth}
\centering \includegraphics[width=3in]{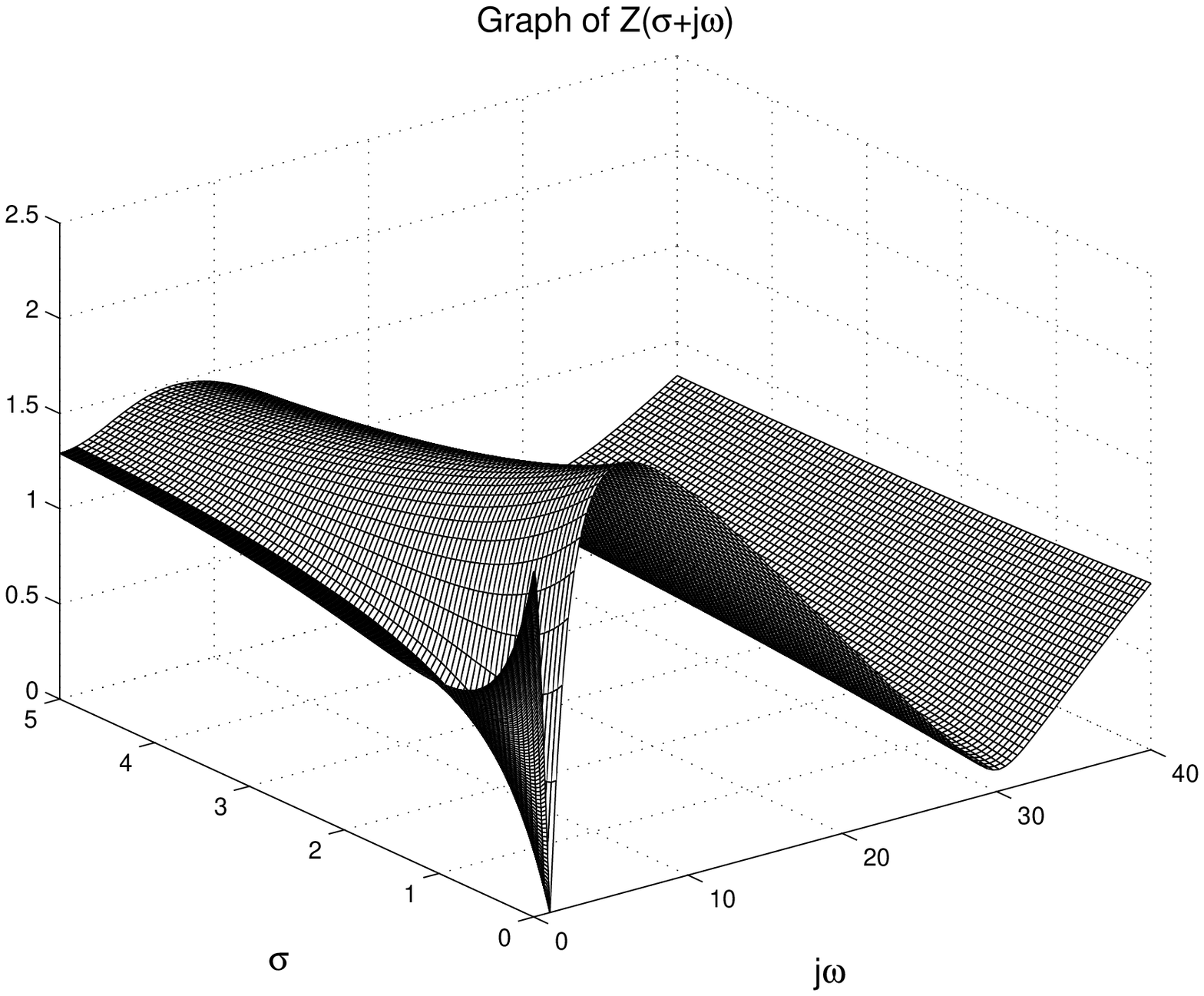}
\caption{$Z(s)$ plot for right half plane}
\label{fig:regionofzeros}
\end{minipage}%
\end{figure}

We will use the design method of the Section 3 to find a stable
suboptimal controller by search for $U$. The central suboptimal
controller ($U=0$) has infinitely many unstable poles as mentioned
before. The algorithm is tried for $u_z=u_p=0$ case, i.e.,
$U(s)=u_\infty$.
\begin{description}
  \item[1)] Fix $\rho=0.814 > \gamma_{opt}=0.8108$,
  \item[2)] $k=-0.9413$ and $f_\infty=1.3567$ are calculated.
  \item[3)] $n_1=1$, $l=0$, $n_1+l$ is odd and $|k|>\frac{1}{f_\infty}$. By using Lemma
  (\ref{eq:inftyvalues}), the admissible values for $u_\infty$ are
  $-0.9909<u_\infty<-0.6668$.
  \item[4)] $L_{1U}(s)$ is stable for $u_\infty\in[-1,0.98]$.
  \item[5)] Overall admissible values for $U$ are
  $u_\infty\in[-0.9909,-0.6668]$. The values of $\omega_{max}$ and $\eta_{max}$ for all admissible $u_\infty$ range can be seen
  in Figure \ref{fig:wetamax}. Since $\eta_{max}$ values do not vary much, the minimum value of $\omega_{max}$ determines
  the optimal $u_{\infty}$ value as $\omega_{max}=19.458$ at $u_{\infty}=-0.813$.
  \item[6)] Figure \ref{fig:regionofzeros} shows the plot of
  $Z(s)=|1+e^{-hs}M(s)F_\rho(s)L_U(s)|$ in the right half plane. The
  function has only right half plane zero at $s=\pm1.056j$, which is
  right half zeros of $E_\rho(s)$. Note that, only one part of right plane is
  graphed since the other half is same.
\end{description}

Therefore, we can conclude that suboptimal controller is stable
for $U(s)=-0.813$ and achieves the $\Hi$ norm $\rho=0.814$.

\subsection{Example}
For given plant $P(s)=e^{-3s}$ and weight functions
$W_1(s)=\left(\frac{2.24+s}{1+s}\right)$ and $W_2(s)=0.5(2.24+s)$,
we can find the optimal performance level as
$\gamma_{opt}=1.9452$. The corresponding optimal $\Hi$ controller
can be written as, \be C_{opt}(s)=E_{\gamma_{opt}}(s)
\frac{F_{\gamma_{opt}}(s)L_{opt}(s)}{1+m_n(s)F_{\gamma_{opt}}(s)L_{opt}(s)}\ee
where \bea
\nonumber m_n(s)&=&e^{-3s}, \\
\nonumber E_{\gamma_{opt}}(s)&=&\frac{1.2162+2.7838s^2}{3.7838(1-s^2)},\\
\nonumber F_\rho(s)&=&5.5119\frac{(1-s)}{\left(2.24+s\right)^2}, \\
\nonumber L_{opt}(s)&=&1. \eea

The optimal controller has unstable poles at $s=0.0292\pm2.2354j
$. Note that since $W_1$ and $W_2$ are proper and improper
respectively, all $\Hi$ controllers will have finitely many
unstable poles by Theorem \ref{eq:finitepoles}. Therefore we can
apply the algorithm in section \ref{eq:finitelymany}.

\begin{figure}
\begin{minipage}[b]{.5\linewidth}
\centering \includegraphics[width=3in]{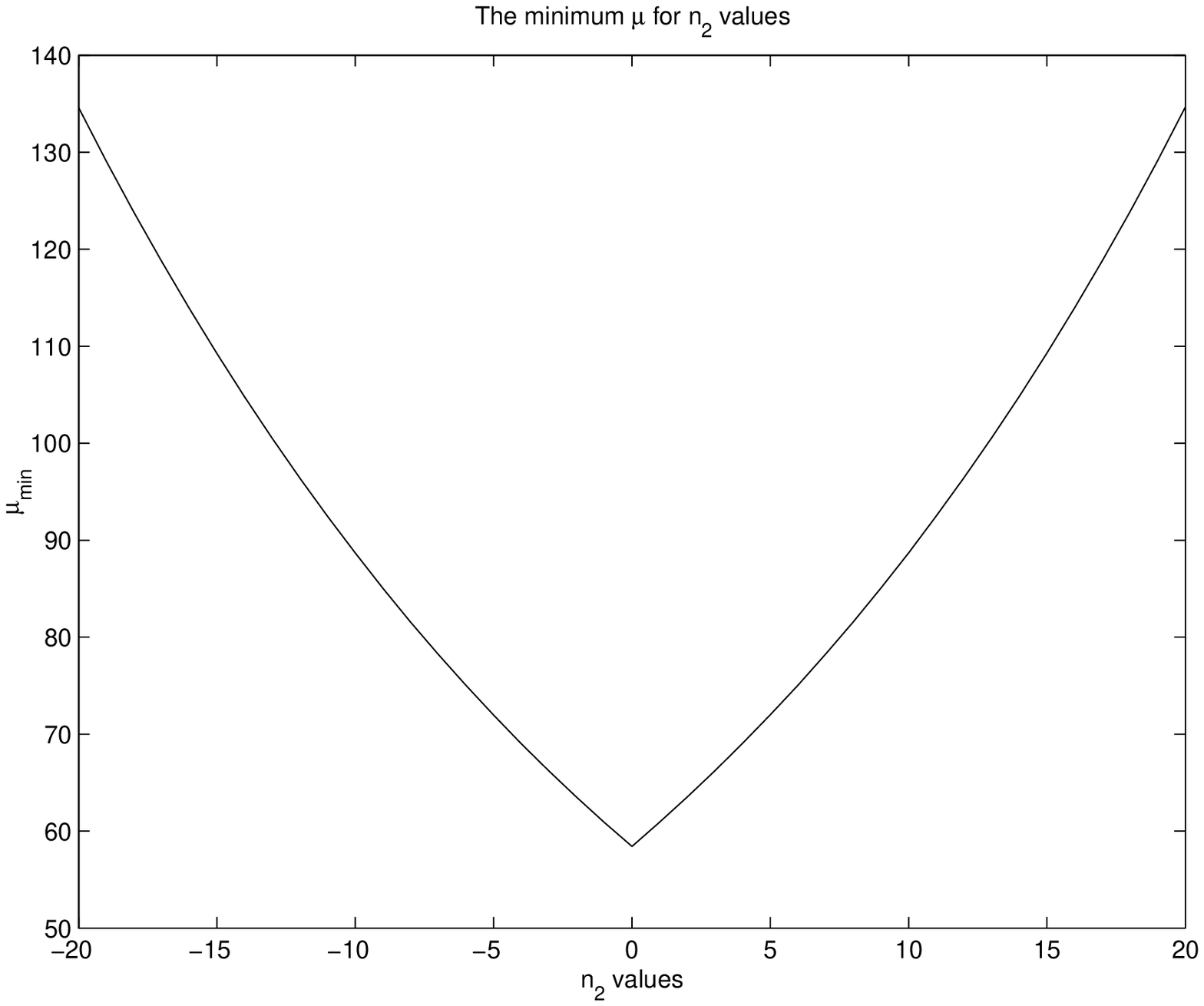}
\caption{$\mu_{min}$ versus $n_2$} \label{fig:musearch}
\end{minipage}%
\begin{minipage}[b]{.5\linewidth}
\centering \includegraphics[width=3in]{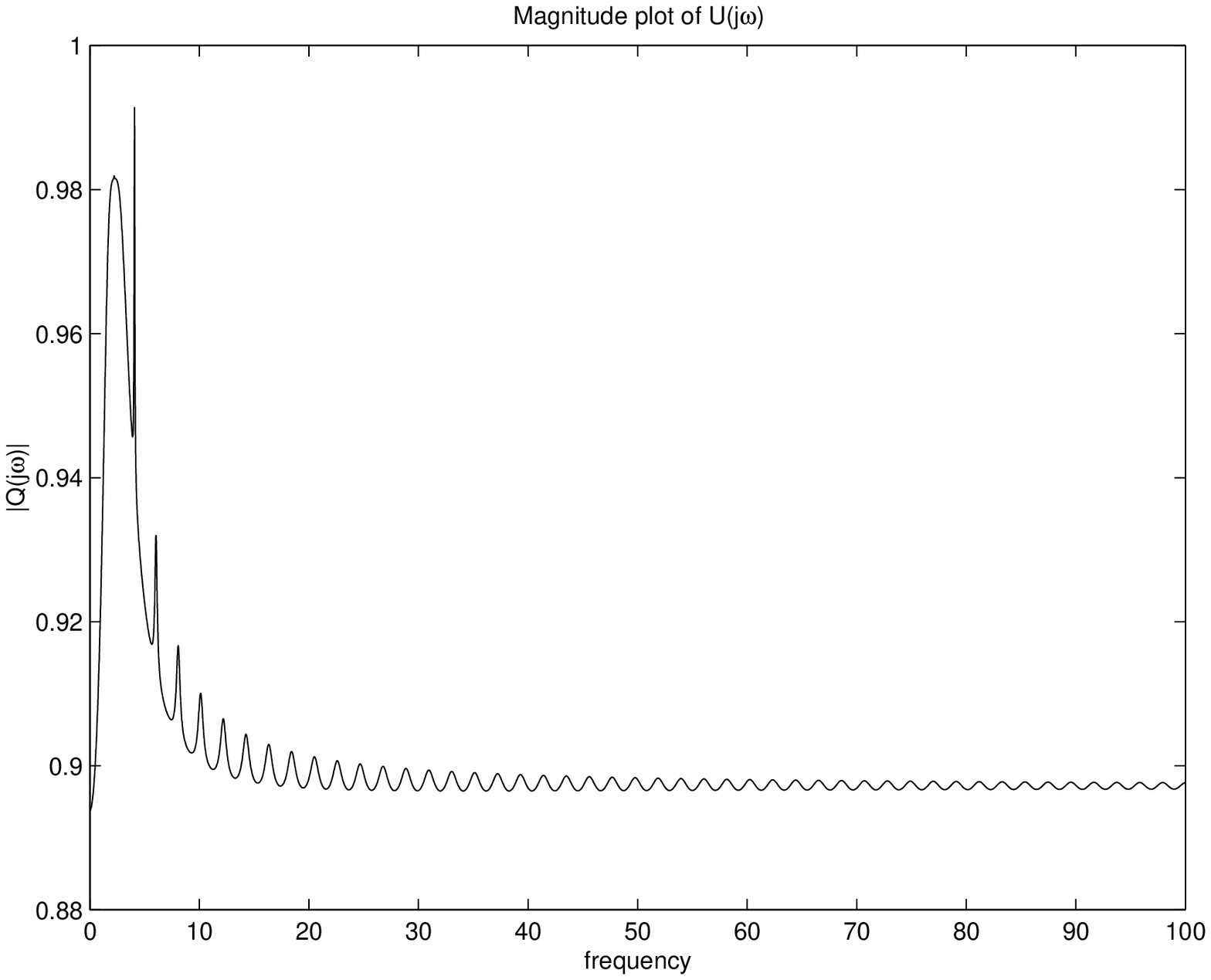}
\caption{Magnitude plot of $U(j\omega)$} \label{fig:Uversusw}
\end{minipage}%
\end{figure}

\begin{figure}
\centering \includegraphics[width=3in]{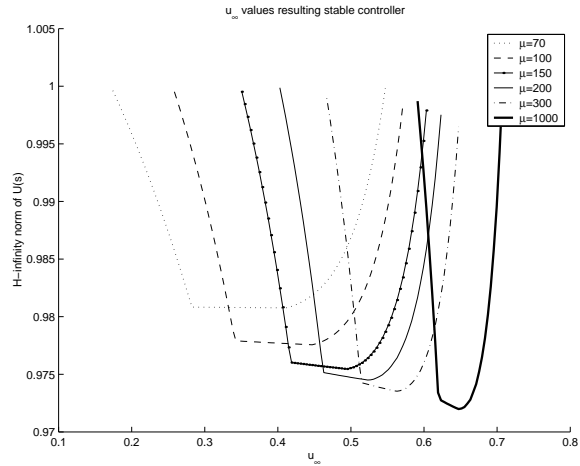}
\caption{$u_\infty$ values resulting stable $\Hi$ controller}
\label{fig:Uinfversusuinf}
\end{figure}

\begin{description}
  \item[1)] Fix $\rho=1.9454>\gamma_{opt}=1.9452$,
  \item[2)] The suboptimal controllers can be written as,
  \be C_{subopt}(s)=E_{\rho}(s)
\frac{F_{\rho}(s)L_U(s)}{1+m_n(s)F_{\rho}(s)L_U(s)}\ee where \bea
\nonumber m_n(s)&=&e^{-3s}, \\
\nonumber E_\rho(s)&=&\frac{1.2154+2.7846s^2}{3.7846(1-s^2)},\\
\nonumber F_\rho(s)&=&5.5115\frac{(1-s)}{(2.24+s)^2}, \\
\nonumber
L_U(s)&=&\frac{L_{2U}(s)}{L_{1U}(s)}=\frac{L_2(s)+L_1(-s)U(s)}{L_1(s)+L_2(-s)U(s)},\\
\nonumber L_2(s)&=&(2.9837+0.9946s),\\
\nonumber L_1(s)&=&(2.9829+s), \eea and $U$ is free parameter such
that $U\in\Hi$, $\|U\|_\infty\leq1$. We can write $P_1$ and $P_2$
as, \bea \nonumber
P_1(s)&=&\frac{L_1(s)+m_n(s)F_\rho(s)L_2(s)}{nE_\rho(s)}, \\
\nonumber
&=&\frac{(2.9829+s)(2.24+s)^2+5.5115(1-s)(2.9837+0.9946s)e^{-3s}}{(1.2154+2.7846s^2)(2.24+s)^2},
\\
\nonumber
P_2(s)&=&\frac{L_2(-s)+m_n(s)F_\rho(s)L_1(-s)}{nE_\rho(s)}, \\
\nonumber
&=&\frac{(2.9837-0.9946s)(2.24+s)^2+5.5115(1-s)(2.9829-s)e^{-3s}}{(1.2154+2.7846s^2)(2.24+s)^2}.
\eea Note that $P_1$ and $P_2$ has unstable zeros at
$0.0287\pm2.2346j$ and $0.0297\pm2.2346j$ respectively. Therefore,
the central controller ($U=0$) for the chosen performance level,
$\rho=1.9458$, is unstable.
  \item[3)] Define the following variables and functions as,
  \bea \nonumber p_i&=&0.0287\pm2.2346j, \quad i=1,2,\\
  \nonumber s_i&=&0.0297\pm2.2346j, \quad i=1,2,\\
  \nonumber
  \tilde{M}_d(s)&=&\frac{(s-p_1)(s-p_2)}{(s+p_1)(s+p_2)}=\frac{s^2-0.0574s+4.9943}{s^2+0.0574s+4.9943},\\
  \nonumber w_i&=&\frac{1}{\tilde{M}_d(s_i)}=58.4002\mp0.7501j,
  \quad i=1,2,\\
  \nonumber z_i&=&\frac{s_i-1}{s_i+1}=0.6598\pm0.7383i
\eea where conformal mapping parameter, $a$, is chosen as $a=1$.
  \item[4)] In order to find the minimum $\mu$ resulting in positive semi-definite Pick
  matrix,\be Q_P\{\mu\}=\left(
\begin{array}{cc}
  \frac{-8.1348+2\ln{\mu}}{0.0196} & \frac{(-8.1348+0.0257j)+2\ln{\mu}+j2\pi(n_2-n_1)}{1.1097-0.9742j} \\
  \frac{(-8.1348-0.0257j)+2\ln{\mu}+j2\pi(n_1-n_2)}{1.1097-0.9742j} & \frac{-8.1348+2\ln{\mu}}{0.0196}
\end{array} \right), \ee we will find the minimum $\mu$ for all possible integer pairs
$(n_1,n_2)$. It is not difficult to do this search since many
integer pairs do not result in positive semi-definite Pick matrix.
For each integer pair, we can find the minimum $\mu$, $\mu_{min}$,
and then $\mu_{opt}$ will be smallest of all $\mu_{min}$. Note
that since Pick matrix depends on difference of integers, we can
normalize the search by taking $n_1=0$. In Figure
\ref{fig:musearch},we can see the minimum $\mu$ values for
integers, $n_2$. The minimum of all $\mu_{min}$ values is
$\mu_{opt}=58.4167$.
  \item[5)] The calculation of $U(s)$ for $\mu_{opt}$ is omitted.
  It does not satisfy the norm condition $\|U\|_\infty\leq1$.
  \item[6)] Fix $\mu=64$ and $n_1=n_2=0$. The interpolation
  conditions for $g(z)$ can be written as,
  \be g(z_i)=0.0915\pm0.0128j, \quad i=1,2. \ee By Nevanlinna-Pick approach, (see e.g.\cite{FOT}),
  \be
  g(z,q)=\frac{(1.0878z^2-1.3782z+0.9804)q(z)+(0.0724z-0.1054)}{(0.9804z^2-1.3782z+1.0878)+(0.0724z-0.1054z^2)q(z)}
  \ee where $q(z)$ is a parameterization term such that $q\in\Hi$ and
  $\|q\|_\infty\leq1$. The search algorithm tries to find $q_o$
  satisfying the norm condition \be \label{eq:normcondition} \left|\left(\frac{1-\mu \tilde{M}_d(j\omega)e^{-G(j\omega,Q)}}{\mu \tilde{M}_d(j\omega)e^{-G(j\omega,Q)}}\right)\left(\frac{P_1(j\omega)}{P_2(j\omega)}\right)\right|\leq1, \quad \forall
  \omega\in[0,\infty)\ee where
  \bea
  \nonumber G(s,Q(s))&=&g\left(\frac{s-1}{s+1},q\left(\frac{s-1}{s+1}\right)\right), \\
  \nonumber          &=&\frac{(0.69s^2-0.2148s+3.4464)Q(s)-(0.0330s^2+0.3556s+0.0330)}{(0.69s^2+0.2148s+3.4464)-(0.0330s^2-0.3556s+0.0330)Q(s)}
  \eea and $Q\in\Hi$, $\|Q\|_\infty\leq1$. We will search for $Q$ satisfying the norm condition (\ref{eq:normcondition}) in the form of
  $Q(s)=u_\infty$ with $|u_\infty|\leq1$. Note that we choose $z_u=p_u=0$ and all
  functions in norm condition, $P_1$, $P_2$, $\tilde{M}_d$, are
  defined before. After search is done, the condition
  (\ref{eq:normcondition}) is satisfied for $u_\infty=0.323$. The
  magnitude of $U(j\omega)$ is smaller than one for all frequency
  values as seen in Figure. (i.e., $\|U\|_\infty=0.9924$). As a
  result, the suboptimal $\Hi$ controller achieving the performance
  level, $\rho=1.9454$, is stable with selection of parameter $U$
  as,
  \be
  U(s)=\left(0.0156\left(\frac{s^2+0.0574s+4.9943}{s^2-0.0574s+4.9943}\right)e^{\left(\frac{0.1899s^2-0.4250s+1.0802}{0.6793s^2+0.3297s+3.4357}\right)}-1\right)\left(\frac{P_1(s)}{P_2(s)}\right).
  \ee
  By the search algorithm, we can find many $u_\infty$ values for
  different $\mu$ resulting in stable $\Hi$ controller at $\rho=1.94584$ provided
  that $U$ satisfies the norm condition for chosen $Q=u_\infty$. The
  various $u_\infty$ values resulting stable $\Hi$ controller can be seen
  in Figure \ref{fig:Uinfversusuinf}. We can observe that as $\mu$
  is increased, the range of $u_\infty$ stabilizing the controller
  decreases and the minimum value of $\|U\|_\infty$ in the $u_\infty$ range becomes
  smaller.
\end{description}

\section{Conclusions}
In this paper, for delay systems, we investigated stability of the
$\Hi$ controllers whose stucture is given in
\cite{TO95},\cite{FOT}. We considered the controllers in two
subsections according to their number of poles (finite, infinite).
For each case, necessary conditions and design methods based on
simple sufficient condition are given to find stable suboptimal
$\Hi$ controllers.

\end{document}